\documentstyle[aps]{revtex}

\begin{document}

\title{Non--standard Construction of Hamiltonian Structures and of
the Hamilton--Jacobi equation.}

\author{Mauricio Herrera $^{1}$ and Sergio A. Hojman $^{2,3}$}

\address{$^{1}$ Departamento de F\'{\i}sica, Facultad de Ciencias F\'{\i}sicas y Matem\'aticas\\
Universidad de Chile, Santiago, Chile\\
$^{2}$ Departamento de F\'{\i}sica, Facultad de Ciencias \\
Universidad de Chile, Casilla 653, Santiago, Chile\\
$^{3}$ Centro de Recursos Educativos Avanzados, CREA\\
Vicente P\'erez Rosales 1356-A, Santiago, Chile}

\maketitle

\begin{abstract}\
Examples of non--standard construction of Hamiltonian structures
for dynamical systems and the respective Hamilton--Jacobi (H--J)
equations, without using Lagrangians, are presented. Alternative
H--J equations for Euler top are explicitly exhibited and solved.
We demonstrate that the stability criterion used by Bridges in
\cite{bridges}, relating the slope of a Casimir function
parametrized by the Lagrange multiplier to critical point type,
depends on the used Hamiltonian structure and it is inadequate
for this reason.
\end{abstract}

\section{Introduction}

The standard construction of Hamiltonian theories starts from a Lagrangian. The
procedure is well known and it is the subject of very many textbooks.
Nevertheless it is interesting to construct hamiltonian structures for
classical systems of differential equations, without using a Lagrangian, which
may even fail to exist, starting from the equation of motion only. These
approaches usually are not systematic, conducting to non unique hamiltonian
structures and deal with non canonical variables (in some systems these
variables are more natural than canonical ones).

One systematic, but complicated procedure to construct hamiltonian structures
using symmetries of classical equations was intoduced by E. Sudarshan and N.
Mukunda in\cite{mukun2}. They also comment the possible relevance of
alternative hamiltonian of classical systems under the quantization procedure. Another
systematic approach due also to Sudarshan and Mukunda \cite{mukun1}, but
developed by J. Marsden, A. Weistein, D. Holm and B. Kupershmidt in \cite{marsd}--\cite{kuper}
is based on definition of Lie -- Poisson structures on duals of Lie algebras.
More recently S. Hojman, et. al \cite{hojman1}--\cite{hojman7} presented a
newly devised method which constitutes a general technique for construction of
hamiltonian structures using symmetries and constants of motion of dynamical systems.

 Non standard introduction of hamiltonian structures enrich the possibilities
we have in the analysis of both classical and quantum systems and for this
reason one must be careful interpreting the results we obtained. In the section
III of the present work we use this fact in the context of analysis of
stability for dynamical systems in the vicinity of a critical point,
demostrating that the criterion of stability used by Bridges in \cite{bridges}
is inadequate.

On the other hand the Hamilton -- Jacobi equation (H--J) is one of the
cornerstones of the theoretical physics. Its role in the theoretical setting of
optics and both classical and quantum mechanic is well established. Its use as
a tool for solving problems is also well known.

The usual way of deriving the H--J equation is based on the lagrangian or
hamiltonian formalisms. In the section IV of this paper we present one way of constructing the
H--J equation starting from equation of motion only, to do this we use the
approaches due to S. Hojman et. al.\cite{hojman1}--\cite{hojman7} for non
standard construction of hamiltonian structures. In some sense this complement
the analysis carry out in the section III, because we only concern here with
the study of regular points of dynamical systems and not concerns singular
points of it, as in the above mentioned section.

With the help of alternative hamiltonian structures introduced by Hojman's
recipe, we define the Poisson brackets on the phase space {\em M} for dynamical
system. Using a Poisson map $\varphi$, {\em M} is reduced to a symplectic
submanifold {\em N}. On {\em N} we can define the $action$ and the respective H--J
equation.

The hamiltonian structures on {\em M} are so on {\em N}. Furthermore the
solutions of dynamical system with initial conditions on {\em N} stay all the
time on it. For these reasons the solution of H--J equation is a particular
solution of dynamical system constrained to {\em N}.

There it is not unique way to define hamiltonian structure on {\em M}, neither
the Poisson map used for reduction of the phase space to submanifold {\em N}
is unique. This means that we can define the action in different ways and
obtain different H--J equations. In this paper we show that solutions obtained
from different H--J equations, from alternative hamiltonian structures on {\em
M} yield correct solutions of dynamical systems. This allow us to speak about
some kind of equivalence between different H--J equations obtained this way.

In the section II, for illustration purpose of using the method introduced by
S. Hojman, we present one example borrowed from D. Armbruster, J.
Guckenheimer, and P. Holmes in \cite{holmes1,holmes2}. This example was
employed in the analysis of normal forms with {\large O}(2) symmetry. Using the
complete set of symmetries of this dynamical system, it is introduced two
alternative hamiltonian structures. With the help of {\large O}(2) symmetry we
can reduce the four equations dynamical system in cartesian coordinates to
three equations in polar coordinates and the dynamical system obtained this
way we will use as one of the example presented in the section II. The physics
motivations for the study of these systems can be found in the above listed
references.

The section III is dedicated to demonstrated that the criterion of Bridges
in \cite{bridges}, that relates the slope of a Casimir function parametrized
by the Lagrange multiplier to critical point type depends of the used hamiltonian
structures and it is inadequate for this reason.

The section IV is devoted to illustrated the construction of H--J in the way
described above. We employ the examples used in the section III and present a
complete analysis of the Euler top.

\section{One example of non--standard construction of hamiltonian structure.}

The contents of this Section illustrates partially the results of
\cite{hojman1}--\cite{hojman7}.

Consider a dynamical system for the complex variables $Z_{1}$ and $Z_{2}$
defined by the following equations:
\begin{eqnarray}
 {\dot Z_{1}}&=&\bar{Z}_{1}Z_{2}\nonumber \\
\mbox{}\label{equ}\\
{\dot Z_{2}}&=&-Z^{2}_{1} \nonumber
\end{eqnarray}

The system (\ref{equ}) can be writing in cartesian coordinates:
\begin{eqnarray}
\dot{X}_{1}&=&X_{1}X_{2}+Y_{1}Y_{2}\nonumber\\
\dot{Y}_{1}&=&X_{1}Y_{2}-Y_{1}X_{2}\nonumber\\
\dot{X}_{2}&=&-X^{2}_{1}+Y^{2}_{1}\label{equc}\\
\dot{Y}_{2}&=&-2X_{1}Y_{1}\nonumber
\end{eqnarray}

Where:
\begin{eqnarray}
Z_{1}&=&X_{1}+iY_{1}\nonumber\\
Z_{2}&=&X_{2}+iY_{2}\nonumber
\end{eqnarray}

A hamiltonian structure for (\ref{equ} or \ref{equc})
  consist of an
antisymmetric matrix, $J^{ab}$ and a Hamiltonian $H$
such that $J^{ab}$ is the Poisson bracket for the dynamical variables, which
are in general non--canonical. In addition to its antisymmetry, the matrix
$J^{ab}$  is required to satisfy Jacobi identity and to reproduce, in
conjunction with the Hamiltonian $H$ the dynamical equation (\ref{equ} or
\ref{equc}) i. e.,
\begin{equation}
{J^{ab}}_{,d} J^{dc} + {J^{bc}}_{,d} J^{da} + {J^{ca}}_{,d}  J^{db}
\equiv
0\,
\label{ji}
\end{equation}
and,
\begin{equation}
J^{ab}  \frac{\partial H}{\partial x^b} = f^a\ .
\label{hami}
\end{equation}

Where:
\begin{equation}  f = \left ( \begin{array}{cc}
X_{1}X_{2}+Y_{1}Y_{2}\\
X_{1}Y_{2}-Y_{1}X_{2}\\
-X^{2}_{1}+Y^{2}_{1}\\
-2X_{1}Y_{1}
\end{array}\right )
\end{equation}

It has been proved \cite{hojman1,hojman2} that one solution to the problem of finding
a Hamiltonian structure for a given dynamical system is provided by one constant
of the motion which may be used as the Hamiltonian $H$, and a symmetry vector
$\eta^{a}$ which allows for the construction of a Poisson matrix $J^{ab}$. The
constant of the motion and the symmetry vector satisfy,
\begin{eqnarray}
{\cal L}_f H &=& 0 \label{cc} \ ,\\
(\partial_t + {\cal L}_f) \eta^a &=& 0  \label{sym} \ ,
\end{eqnarray}

respectively, where ${\cal L}_f$ is the Lie derivative along $f$ (for a
definition,
see \cite{wein}, for instance).
In addition, it is required that the deformation $K$ of $H$ along $\eta^{a}$,
\begin{equation}
K \equiv \frac{\partial H}{\partial x^a}\eta^a = {\cal L}_\eta H
 \ ,
\label{deform}
\end{equation}
be non-vanishing.

The Poisson matrix $J^{ab}$ is constructed as the antisymmetrized product
of the flow vector $f^{a}$ and the ``normalized'' symmetry vector
$\eta^{b}/K$,
\begin{equation}
J^{ab} =\frac{1}{K}(f^a \eta^b - f^b \eta^a)  \ .
\label{ans}
\end{equation}

The Poisson matrix so constructed has rank 2 and it is, therefore, singular.
  Adding together two Poisson matrices constructed according
to (\ref{ans}) will not increase its rank. It will just redefine the
symmetry vector used to construct it. One method to increase the rank
of such a Poisson matrix is presented in \cite{hojman1}.

The system (\ref{equ}) is invariant under the transformation:
\begin{eqnarray}
Z_{1}&\rightarrow& Z_{1}\exp(i\Phi)\nonumber \\
\mbox{} \label{trf}\\
Z_{2}&\rightarrow& Z_{2}\exp(2i\Phi)\nonumber
\end{eqnarray}

In cartesian coordinates the transformation (\ref{trf}) is writing as:
\[ \left( \begin{array}{cc}
\tilde{\cal X}_{1}\\
\tilde{\cal X}_{2}
\end{array} \right)= \left ( \begin{array}{cc}
\tilde{X}_{1}\\
\tilde{Y}_{1}\\
\tilde{X}_{2}\\
\tilde{Y}_{2}
\end{array} \right ) =
\left( \begin{array}{cc}
{\cal R}(\Phi) & 0\\
0 & {\cal R}(2\Phi)
\end{array}\right)\]

Where:
\[ {\cal R}(\Phi)= \left( \begin{array}{cc}
\cos \Phi & -\sin \Phi \\
\sin \Phi  & \cos \Phi
\end{array}\right)\]

From this, we have the symmetry vector:

 \begin{equation}  \eta^{(1)} = \left ( \begin{array}{cc}
-Y_{1}\\
X_{1}\\
-2Y_{2}\\
2X_{2}
\end{array}\right )
\end{equation}

Another symmetry transformation, involving space coordinates
and time of (\ref{equc}) is:
 \[ \left( \begin{array}{cc}
\tilde{\cal X}_{1}\\
\tilde{\cal X}_{2}\\
\tilde{t}
\end{array} \right)= \left ( \begin{array}{cc}
{\cal X}_{1}\\
{\cal X}_{2}\\
\lambda^{-1}t
\end{array} \right ) \]

The infinitesimal version of this transformation is:

 \begin{equation} \tilde{\eta}^{(2)} = \left ( \begin{array}{cc}
{\cal X}_{1}\\
{\cal X}_{2}\\
-t
\end{array}\right )
\end{equation}

This vector we can equivalently write in the form: (see\cite{hojman1,hojman2})
\begin{equation} \eta^{(2)} = \left ( \begin{array}{cc}
X_{1}+t(X_{1}X_{2}+Y_{1}Y_{2})\\
Y_{1}+t(X_{1}Y_{2}-Y_{1}X_{2})\\
X_{2}+t(-X^{2}_{1}+Y^{2}_{1})\\
Y_{2}-t(2X_{1}Y_{1})
\end{array}\right )
\end{equation}

The symmetry vectors $\eta^{(1)}$, $\eta^{(2)}$ and $f$ satisfy the symmetry
equation (\ref{sym}) and the Lie algebra:
\begin{eqnarray}
 [\eta^{(2)}, f]&=&-f\nonumber \\
 {[}\eta^{(1)}, \eta^{(2)}]&=&0 \\
 {[}\eta^{(1)}, f]&=&0\nonumber
\end{eqnarray}

For the system (\ref{equ}) we have the following constants of motion:
\begin{eqnarray}
 C_{1}&=&|Z_{1}|^2+ |Z_{2}|^2 \nonumber \\
C_{2}&=&\frac{1}{2i}(Z^{2}_{1}\bar{Z}_{2}-\bar{Z}^{2}_{1}Z_{2}) \nonumber
\end{eqnarray}

Or in cartesian coordinates:
\begin{eqnarray}
C_{1}&=&X^{2}_{1}+Y^{2}_{1}+X^{2}\_{2}+Y^{2}_{2}\\
C_{2}&=&2X_{1}Y_{1}X_{2}-Y_{2}(X^{2}_{1}-Y^{2}_{1})
\end{eqnarray}

We can now construct some alternative Poisson structures taking as Hamiltonian
$H$ = $C_{1}$ and the Poisson matrix:
\begin{equation}
J^{ab} =\frac{1}{K}(f^a \eta^{(2)b} - f^b \eta^{(2)a})  \ .
\end{equation}

Where $K$ is the deformation of $H$ = $C_{1}$ along $\eta^{(2)}$,
\begin{equation}
K \equiv \frac{\partial H}{\partial x^a}\eta^a = {\cal L}_\eta H = 2C_{1}
 \ ,
\end{equation}

Or taking $C_{2}$ as a Hamiltonian $H$, and Poisson matrix:
 \begin{equation}
J^{ab} =\frac{1}{Q}(f^a \eta^{(2)b} - f^b \eta^{(2)a})  \ .
\end{equation}

Where $Q$ is the deformation of $H$ = $C_{2}$ along $\eta^{(2)}$,
\begin{equation}
Q \equiv \frac{\partial H}{\partial x^a}\eta^a = {\cal L}_\eta H = 3C_{2}
 \ ,
\end{equation}

The deformations of both $C_{1}$ and $C_{2}$ along $\eta^{(1)}$ are vanishing
and for this reason we can not use this vector in the construction of Poisson
matrix with the above procedure.

Finally we note that introducing polar coordinates and using the {\large O}(2)
 symmetry, the system (\ref{equ}) can
be writing as:
\begin{eqnarray}
\dot{r}_{1}&=&r_{1}r_{2}\cos \Theta \nonumber\\
\dot{r}_{2}&=&\pm r^{2}_{1} \cos \Theta \label{equp}\\
\dot{\Theta}&=&- \left (2 r_{2}\pm \frac{r^{2}_{1}}{r_{2}} \right ) \sin
\Theta \nonumber
\end{eqnarray}

Where:
$ Z_{j}=r_{j}\exp \Phi_{j}$,  $\Theta = 2\Phi_{1}-\Phi_{2}$

\section{Hamiltonian structures and linear stability criterion.}
In this section we use another way to construct Poisson structures based on the knowledge of
constants of motion for the dynamical system under consideration (See
\cite{hojman2} for detail) to demonstrate that stability criterion used by Bridges in
\cite{bridges} strongly depends of the used Poisson structures.
Let us consider as the first example the one we presented in the preceding section.
\smallskip

\underline {Example I}:
\begin{eqnarray}
\dot{X}_{1}&=&X_{1}X_{2}\cos \Phi \nonumber\\
\dot{X}_{2}&=&\pm X^{2}_{1} \cos \Phi \label{ex1}\\
\dot{\Phi}&=&- \left (2 X_{2}\pm \frac{X^{2}_{1}}{X_{2}} \right ) \sin
\Theta \nonumber
\end{eqnarray}

The phase space of (\ref{ex1}) is:
\[M={\large \{ }\Phi, X_{1},X_{2}:(X_{1},X_{2})
\in {\cal R}^{2+},0\leq \Phi \leq 2\pi{\large \}} \]

The integrals of motion for this system are:
\begin{eqnarray}
C_{1}&=&\frac{1}{2}(X^{2}_{1}\mp X^{2}_{1})\\
C_{2}&=&X^{2}_{1}X_{2} \sin \Phi
\end{eqnarray}

Taking as a Hamiltonian $H=C_{1}$ and the function $\Psi=C_{2}$ we have
the following Poisson structure:
\medskip

\underline{I Poisson Structure}:
\[ J^{(1)ab}=\frac{1}{X_{1}X_{2}}\varepsilon^{abc}\frac{\partial \Psi}{\partial X^{c}}
\]
 \begin{equation}\|J^{(1)ab}\|=
\left ( \begin{array}{lcr}
0&X_{1}\cos \Phi&-\frac{X_{1}}{X_{2}}\sin \Phi\\
-X_{1}\cos \Phi&0&2\sin \Phi\\
\frac{X_{1}}{X_{2}}\sin \Phi&-2\sin \Phi&0
\end{array}\right )
\label{pois1}
\end{equation}

It is not difficult to see that (\ref{pois1}) together with $H=C_{1}$ provide a
Hamiltonian formulation for (\ref{ex1}), and that $\Psi$ is the Casimir function
( i. e. function which has  vanishing Poisson bracket relations with any other
dynamical quantity. For detail, see  \cite{little,mukun1} for instance) for this
 Poisson structure.

In other words we see that,
\[
\dot{{\cal X}}=\{{\cal X},H\}_{M}  \]

Where:
${\cal X}=(X_{1},X_{2},\Phi)$

Let $\lambda $ be a Lagrange multiplier, then a necessary condition for
$\cal X$ to be a critical point of (\ref{ex1}) given a Hamiltonian
$H$ and a Casimir $\Psi$ is that:
\[ \nabla H({\cal X})-\lambda \nabla\Psi({\cal X})=0 \]
Or
\[ \left( \begin{array}{cc}
X_{1}\\
X_{2}\\
0
\end{array} \right)- \lambda \left ( \begin{array}{cc}
2X_{1}X_{2}\sin\Phi\\
X^{2}_{1}\sin\Phi\\
X^{2}_{1}X_{2}\cos\Phi
\end{array} \right ) =0\]

From this equations we have the critical point:
\begin{equation}
P_{c}= \left (\frac{1}{\sqrt{2}\lambda},\frac{1}{2\lambda},\frac{\pi}{2}\right)
\label{critical1}
\end{equation}

On the other hand, the derivative of $\Psi_{P_{c}}$ and $H_{P_{c}}$ with
respect to Lagrange multiplier are:
\begin{eqnarray}
\frac{d\Psi_{P_{c}}}{d\lambda }&=&-\frac{3}{4\lambda^4} \label{casimir1}\\
\frac{d H_{P_{c}}}{d\lambda}&=&-\frac{3}{4\lambda^3}
\end{eqnarray}

Liniarizing the Poisson system (\ref{ex1}) about a critical point $P_{c}$:
${\cal X}^{i}=P_{c}+\xi^{i}$, satisfying $\nabla H=\lambda\nabla\Psi$ we have:
\begin{equation}
\dot{\xi}^{i}=\frac{\partial}{\partial X^{m}}\left
[\frac{1}{X_{1}X_{2}}(\nabla\times\nabla H)^{i}\right]_{P{c}}\xi^{m}=
\left[\frac{1}{X_{1}X_{2}}\nabla\Psi\times(D^{2}H-\lambda
D^2\Psi)_{P_{c}}\right]^{i}_{m}\xi^{m}
\end{equation}

Where:
\begin{equation}
\frac{1}{X_{1}X_{2}}\nabla\Psi\times(D^{2}H-\lambda D^2\Psi)_{P_{c}}=
\left ( \begin{array}{lcr}
0&0&-\frac{1}{2\sqrt{2}\lambda^2}\\
0&0&\frac{1}{2\lambda^{2}}\\
2\sqrt{2}&-4&0
\end{array}\right )
\label{matr1}
\end{equation}

One eigenvalue of (\ref{matr1}) is zero and the other two are given by:
\begin{equation}
\mu^{2}=-\frac{3}{\lambda^2}=4\lambda^2\frac{d\Psi_{P_{c}}}{d\lambda}
\end{equation}

Where we have used (\ref{casimir1}).

This prove that for the I Poisson structure of dynamical system (\ref{ex1})
the critical point is elliptic (purely imaginary eigenvalues) when $d\Psi
/d\lambda <0$ and hyperbolic when $d\Psi/d\lambda >0$. This statement is agree
with the stability criterion used by Bridges in \cite{bridges}. But this
criterion depends of the used Poisson structure as we will see, taking
another Poisson structure the criterion change.

Another Poisson structure for (\ref{ex1}) is obtained taking $C_{2}$ as a
Hamiltonian $H$ and $C_{1}$ as a Casimir $\Psi$.
\medskip

\underline{II Poisson Structure}:
\[ J^{(2)ab}=-\frac{1}{X_{1}X_{2}}\varepsilon^{abc}\frac{\partial \Psi}{\partial X^{c}}
\]
 \begin{equation}\|J^{(2)ab}\|=
\left ( \begin{array}{lcr}
0&0&\frac{1}{X_{1}}\\
0&0&\pm\frac{1}{X_{2}}\\
-\frac{1}{X_{1}}&\mp\frac{1}{X_{2}}&0
\end{array}\right )
\label{pois2}
\end{equation}

The critical point of (\ref{ex1}) using the second Poisson structure is
obtained from the equation:
\[ \nabla H({\cal X})-\beta \nabla\Psi({\cal X})=0 \]

From this, we have:
\begin{equation}
P_{c}= \left (\frac{\beta}{\sqrt{2}},\frac{\beta}{2},\frac{\pi}{2}\right)
\label{critical2}
\end{equation}

Note that directly from the equation of motion (\ref{ex1}) the critical point
must satisfy the relation: $X^{2}_{1}=2X^{2}_{2}, \Phi=\frac{\pi}{2}$.

The derivative of $\Psi_{P_{c}}$ and $H_{P_{c}}$ with
respect to Lagrange multiplier are:
\begin{eqnarray}
\frac{d\Psi_{P_{c}}}{d\beta }&=&\frac{3}{4}\beta \label{casimir2}\\
\frac{d H_{P_{c}}}{d\beta}&=&\frac{3}{4}\beta^{2}
\end{eqnarray}
Liniarizing the Poisson system (\ref{ex1}) about a critical point $P_{c}$ we
have the matrix:
\begin{equation}
-\frac{1}{X_{1}X_{2}}\nabla\Psi\times(D^{2}H-\beta D^2\Psi)_{P_{c}}=
\left ( \begin{array}{lcr}
0&0&-\frac{\beta^{2}}{2\sqrt{2}}\\
0&0&\frac{\beta^2}{2}\\
2\sqrt{2}&-4&0
\end{array}\right )
\label{matr2}
\end{equation}

The eigenvalues of (\ref{matr2}) are zero and:
\begin{equation}
\mu^{2}=-3\beta^2=-4\beta\frac{d\Psi_{P_{c}}}{d\beta}
\end{equation}

Where we have used (\ref{casimir2}).
This prove that for the II Poisson structure of dynamical system (\ref{ex1})
the critical point is elliptic (purely imaginary eigenvalues) when $d\Psi
/d\beta >0$ and hyperbolic when $d\Psi/d\beta <0$. This statement is
completly opposite to the mentioned above stability criterion used by Bridges
in \cite{bridges}. This fact confirms that this criterion depends of the used
 Poisson structure.

As the second example illustrating this fact we use the system presented by
T. Bridges in \cite{bridges}.
\smallskip

\underline {Example II}:
\begin{eqnarray}
\dot{X}_{1}&=&2X_{3} \nonumber\\
\dot{X}_{2}&=&2Q(X_{1})X_{3} \label{ex2}\\
\dot{X}_{3}&=&X_{2}+Q(X_{1})X_{1} \nonumber
\end{eqnarray}

Where $Q(X_{1})$ is a real polynomial in $X_{1}$.
The phase space of (\ref{ex2}) is:
$M= {\cal R}^{3+}$

The integrals of motion for this system are:
\begin{eqnarray}
C_{1}&=&X^{2}_{3}-X_{1}X_{2}\\
C_{2}&=&X_{2}-\int_{0}^{X_{1}}Q(s) ds
\end{eqnarray}
\medskip

\underline{I Poisson Structure} $H=C_{1}$, and $\Psi=C_{2}$.
\[ J^{(1)ab}=\varepsilon^{abc}\frac{\partial \Psi}{\partial X^{c}}
\]

 \begin{equation}\|J^{(1)ab}\|=
\left ( \begin{array}{lcr}
0&0&1\\
0&0&Q(X_{1})\\
-1&-Q(X_{1})&0
\end{array}\right )
\label{pois12}
\end{equation}

With this Poisson structure the equation of motion (\ref{ex2}) take the simple
form: $\dot{\vec{X}}=\nabla\Psi\times\nabla H$.
When $\nabla\Psi$ and $\nabla H$ are parallel the system (\ref{ex2}) has a
critical point.
\[ \nabla H({\cal X})+\lambda\nabla\Psi({\cal X})=0 \]

From this, we have:
\begin{equation}
P_{c}= (\lambda,-\lambda Q(\lambda),0)
\end{equation}

The derivative of $\Psi_{P_{c}}$ and $H_{P_{c}}$ with
respect to Lagrange multiplier are:
\begin{eqnarray}
\frac{d H_{P_{c}}}{d\lambda }&=&2\lambda Q(\lambda)+\lambda^2\frac{d Q}{d\lambda} \label{casimir12}\\
\frac{d\Phi_{P_{c}}}{d\lambda}&=&-\left[Q(\lambda)+\lambda
\frac{d Q}{d\lambda}\right ]\label{casimir21}
\end{eqnarray}

Liniarizing the Poisson system (\ref{ex2}) about a critical point $P_{c}$:
$X^{i}=P_{c}+\xi^{i}$, we have:
\begin{equation}
\dot{\vec{\xi}}=[\nabla\Psi\times(D^{2}H+\lambda D^2\Psi)_{P_{c}}]\vec{\xi}
\end{equation}

Where:
\begin{equation}
\nabla\Psi\times(D^{2}H-\beta D^2\Psi)_{P_{c}}=
\left ( \begin{array}{lcr}
0&0&2\\
0&0&2Q(\lambda)\\
\lambda\frac{d Q}{d\lambda}+Q(\lambda)&1&0
\end{array}\right )
\label{matr21}
\end{equation}
The eigenvalues of (\ref{matr21}) are zero and:
\begin{equation}
\mu^{2}=2\left(2Q+\lambda\frac{d Q}{d\lambda} \right)=-2\frac{d\Psi_{P_{c}}}{d\lambda}
\end{equation}

Where we have used (\ref{casimir21}).
We have proved that for the I Poisson Structure of dynamical system
(\ref{ex2}) the critical point is elliptic when $d\Psi
/d\lambda >0$ and hyperbolic when $d\Psi/d\beta <0$.
\medskip

\underline{II Poisson Structure} $H=C_{2}$, and $\Psi=C_{1}$.
\[ J^{(2)ab}=\varepsilon^{abc}\frac{\partial \Psi}{\partial X^{c}}
\]

 \begin{equation}\|J^{(2)ab}\|=
\left ( \begin{array}{lcr}
0&2X_{3}&X_{1}\\
-2X_{3}&0&-X_{2}\\
-X_{1}&X_{2}&0
\end{array}\right )
\label{pois22}
\end{equation}

With the help of the Poisson structure (\ref{pois22}) T. Bridges in \cite{bridges}
obtained  the opposite criterion to the one proved above. This means that we
can not use the  slope of a Casimir function parametrized
by the Lagrange multiplier to establish the critical point type, independently
 of the used hamiltonian structure.

\section{The Hamilton--Jacobi Equations}

Consider the following equation of motion
\begin{equation}
\frac{dx^a}{dt}=f^a(x^b) \,\, \,\,\,\,\,\,\,\\\ a,b=1...N
\label{eqmo}
\end{equation}

If we have a hamiltonian structure on the phase space $M$ of (\ref{eqmo}) with
a Casimir function $\Psi$, then the Casimir determine the geometry of the
phase space. $\Psi$ is, of course, a constant of the motion of dynamical
system (\ref{eqmo}). The equation $\Psi(x^b)=C$ for different values of $C$ define a
foliation of the phase space $M$. Each leaf of the foliation is an invariant
submanifold, furthermore it is a symplectic submanifold $N$ of the phase
space $M$. Using a Poisson map $\varphi$, i. e. a transformation
$\varphi : M \rightarrow N$ that conserve the Poisson bracket:
$\{F(\varphi),H(\varphi)\}_{M}=\{F,G\}_{N}(\varphi)$, for any functions
$F,H:N\rightarrow {\cal R}$, we can define canonical variables,
 and an $action$ on $N$. The Hamilton--Jacobi (H--J) equation for this
action yields a particular solution of dynamical system constrained to {\em N}.
Let's illustrate the above procedure using the first example (\ref{ex1}) of
the previous section with the II Poisson structure (\ref{pois2}). For this
srtucture we have the following Poisson brackets:
\begin{eqnarray}
\{X_{1},X_{2}\}_{M}=0 \\
\{X_{1},\Phi\}_{M}=\frac{1}{X_{1}}\label{hj1}\\
\{X_{2},\Phi\}_{M}=\pm\frac{1}{X_{2}}
\end{eqnarray}

From (\ref{hj1}) we have:
\begin{equation}
X_{1}\{X_{1},\Phi\}_{M}=\{X_{1},X_{1}\Phi\}=1\nonumber
\end{equation}

This suggest the Poisson map $\varphi : M \rightarrow N$ , given by the inverse of:
\begin{equation}
X_{1}=q,\, \Phi =\frac{p}{q},\,X_{2}=\sqrt{2C_{1}\pm q^2}\label{poismap}
\end{equation}

Where $N$ is the submanifold defined by: $\Psi(X_{1},X_{2},\Phi)=C_{1}$

We can show that $\varphi$ is a Poisson map, moreover it define canonical brackets
on $N$:
\begin{eqnarray}
\{X_{1},X_{1}\Phi\}_{M}=\{q,\frac{pq}{q}\}_{N}=\{q,p\}_{N}=1\nonumber\\
\{X_{1},X_{2}\}_{M}=\{q,\sqrt{2C_{1}\pm q^2}\}_{N}=0\nonumber\\
\{X_{2},\Phi\}_{M}=\{\sqrt{2C_{1}\pm
q^2},\frac{p}{q}\}_{N}=\pm\frac{1}{\sqrt{2C_{1}\pm q^2}}=\pm\frac{1}{X_{2}}\nonumber
\end{eqnarray}

On $N$ in $(p,q)$ coordinates the Hamiltonian $H$ take the form:
\begin{equation}
H=q^2\sqrt{2C_{1}\pm q^2}\sin\frac{p}{q}
\end{equation}

Using this Hamiltonian and introducing the action by the transformation:
$q\rightarrow q$ and $p\rightarrow\frac{\partial S}{\partial q}$.
We have the H--J:

\begin{equation}
\frac{\partial S}{\partial t}+q^2\sqrt{2C_{1}\pm
q^2}\sin\frac{1}{q}\frac{\partial S}{\partial q}=0
\label{H--J1}
\end{equation}

Taking the solution of (\ref{H--J1}) in the form:
\[S=Q(q)-Et\]

Replacing in the H--J equation we find:
\begin{equation}
t-t_{0}=\frac{\partial Q}{\partial E}=\int^{q}_{0}\frac{\acute{q}d\acute{q}}
{\sqrt{\acute{q}^4(2C_{1}\pm\acute{q}^2)-E^2}}
\label{elliptic}
\end{equation}

The integral (\ref{elliptic}) is an elliptic integral and has solution in
terms of special functions (The cnoidal and snoidal functions). The solutions
$q(t)$ and from the Poisson map (\ref{poismap}), the solutions of dynamical
system (\ref{ex1}) lie on concentric closed curves constrained
to the submanifold $N$. We can deduce immediately that all solutions of
(\ref{ex1}) are periodic except for the stationary point (\ref{critical2}).

Let's take now the second example (\ref{ex2}) of the previous section, with
the Poisson structure (\ref{pois12}). We have the following Poisson bracket on
the phase space $M$:
\begin{eqnarray}
\{X_{1},X_{3}\}_{M}=1 \\
\{X_{2},X_{3}\}_{M}=Q(X_{1})\label{hj2}\\
\{X_{1},X_{2}\}_{M}=0
\end{eqnarray}

Using the Poisson map $\varphi : M \rightarrow N$ , given by the inverse of:
\begin{equation}
X_{1}=q,\, X_{3} =p,\,X_{2}=C_{1}+\int^{q}_{0}Q(s)ds\label{poismap1}
\end{equation}

where $N$ is the submanifold defined by: $\Psi(X_{1},X_{2},X_{3})=C_{1}$,
we obtain,  the canonical brackets on $N$ with the Hamiltonian:
\begin{equation}
H=p^2-q\left( C_{1}+\int^{q}_{0}Q(s)ds\right)
\end{equation}

Introducing the action by the transformation:
$q\rightarrow q$ and $p\rightarrow\frac{\partial S}{\partial q}$.
We have the H--J equation:

\begin{equation}
\frac{\partial S}{\partial t}+\left(\frac{\partial S}{\partial q}\right)^{2}
-q\left( C_{1}+\int^{q}_{0}Q(s)ds\right) =0
\label{H--J2}
\end{equation}

Taking the solution of (\ref{H--J2}) in the form:
\[S=K(q)-Et\]

Replacing in the H--J equation we find:
\begin{equation}
2(t-t_{0})=\frac{\partial K}{\partial E}=\int^{q}_{0}\frac{d\acute{q}}
{\sqrt{\acute{q}\left( C_{1}+\int^{\acute{q}}_{0}Q(s)ds\right))+E}}
\label{elliptic1}
\end{equation}

Taking as an example $Q(s)=1$ we find the following solution for (\ref{ex2}):
\begin{eqnarray}
X_{1}&=&\frac{\left[Ae^{2(t-t_{0})}-\frac{C_{1}}{2}\right]^{2}-E}
{Ae^{2(t-t_{0})}}\nonumber\\
X_{2}&=&\frac{\left[Ae^{2(t-t_{0})}+\frac{C_{1}}{2}\right]^{2}-E}
{Ae^{2(t-t_{0})}} \label{solution1}\\
X_{3}&=&\frac{\left[Ae^{2(t-t_{0})}-\frac{C_{1}}{2}\right]
\left[Ae^{2(t-t_{0})}+\frac{C_{1}}{2}\right]}
{2Ae^{2(t-t_{0})}}\nonumber
\end{eqnarray}

Where the solution is obtained for:
$E-\frac{C^{2}_{1}}{4}\geq 0$ and $A=\sqrt{E}+\frac{C_{1}}{2}$
\smallskip
For other expressions of the polynomial $Q(s)$, the solutions of (\ref{ex2})
are expressed in terms of elliptic functions.

Alternatively we can use, with this example the Poisson structure
(\ref{pois22}). For this structure we have the following Poisson brackets on $M$:
 \begin{eqnarray}
\{X_{1},X_{2}\}_{M}=2X_{3} \\
\{X_{1},X_{3}\}_{M}=X_{1}\label{hj3}\\
\{X_{3},X_{2}\}_{M}=X_{3}
\end{eqnarray}

Using the Poisson map $\varphi : M \rightarrow N$ , given by the inverse of:
\begin{equation}
X_{1}=e^{-q},\, X_{3} =p,\,X_{2}=e^{q}p^{2}\label{poismap2}
\end{equation}
where $N$ is the submanifold defined by: $\Psi(X_{1},X_{2},X_{3})=C_{2}$,
we obtain the canonical brackets on $N$ with the Hamiltonian:
\begin{equation}
H=qp^2-\int^{e^{-q}}_{0}Q(s)ds
\end{equation}

The obtained H--J equation is:
\begin{equation}
\frac{\partial S}{\partial t}+e^{q}\left(\frac{\partial S}{\partial q}\right)^{2}
-\int^{e^{-q}}_{0}Q(s)ds =0
\label{H--J3}
\end{equation}

From (\ref{H--J3}) we find an implicit equation for $q(t)$:
\begin{equation}
2(t-\tilde{t}_{0})=\int^{q}_{0}\frac{d\acute{q}}
{\sqrt{e^{-\acute{q}}\left( \int^{e^{-\acute{q}}}_{0}Q(s)ds+E\right)}}
\label{elliptic2}
\end{equation}

 and from $q(t)$ we
obtain a solution of (\ref{ex2}).

Finally we present a similar analysis for the Euler top. The free body
equations of motion are:
\begin{equation}
\dot{\vec{L}}=\vec{\Omega}\times\vec{L},\,\,\,\Omega=\frac{\vec{L}}{I}
\label{euler}
\end{equation}

Where $I=$diag$(I_{1},I_{2},I_{3})$ is the diagonalized moment of inertia tensor,
$I_{1},I_{2},I_{3}>0$. $\vec{L}$ is the angular momentum, and $\vec{\Omega}$
is the angular velocity.

The dynamical system (\ref{euler}) has the following conserved quantities:
\begin{eqnarray}
C_{1}&=&\frac{1}{2}\left(\frac{L^{2}_{1}}{I_{1}}+\frac{L^{2}_{2}}{I_{2}}+\frac{L^{2}_{3}}{I_{3}}\right) \\
C_{2}&=&L^{2}_{1}+L^{2}_{2}+L^{2}_{3}
\end{eqnarray}

This system is Hamiltonian with $H=C_{1}$, $\Psi=C_{2}$ and the following
Poisson matrix:
\[ J^{(1)ab}=\frac{1}{2}\varepsilon^{abc}\frac{\partial \Psi}{\partial X^{c}}
\]
 \begin{equation}\|J^{(1)ab}\|=
\left ( \begin{array}{lcr}
0&L_{3}&-L_{2}\\
-L_{3}&0&L_{1}\\
L_{2}&-L_{1}&0
\end{array}\right )
\label{euler1}
\end{equation}

The phase space of (\ref{euler}) is $M={\cal R}^3$. $M$ is a Poisson manifold
with the brackets:
 \begin{eqnarray}
\{L_{1},L_{2}\}_{M}=L_{3}\nonumber \\
\{L_{2},L_{3}\}_{M}=L_{1}\label{bracket1}\\
\{L_{3},L_{1}\}_{M}=L_{2}\nonumber
\end{eqnarray}

Mapping $\varphi:M\rightarrow N$, with the Poisson map defined by:
\begin{eqnarray}
L_{1}&=&\sqrt{\lambda -p^2}\cos q\nonumber\\
L_{2}&=&\sqrt{\lambda -p^2}\sin q\label{eulerpoi1}\\
L_{3}&=&p\nonumber
\end{eqnarray}

and $N$ being a submanifold defined by the equation: $\Psi
(L_{1},L_{2},L_{3})=C_{2}$. We obtain the canonical brackets
from (\ref{bracket1}), for the variables $p$ and $q$ on the symplectic
submanifold $N$.

Introducing the action by the transformation:
$q\rightarrow q$ and $p\rightarrow\frac{\partial S}{\partial q}$.
We have the H--J equation:
\begin{equation}
\frac{\partial S}{\partial t}+\frac{1}{2I_{1}}\left[\lambda-\left(\frac{\partial
S}{\partial q}\right)^2\right]\cos^{2}q+\frac{1}{2I_{2}}\left[\lambda-\left(\frac{\partial
S}{\partial q}\right)^2\right]\sin^{2}q+\frac{1}{2I_{3}}\left(\frac{\partial
S}{\partial q}\right)^2=0
\label{HJeuler}
\end{equation}

Taking the solution of (\ref{HJeuler}) in the form:
\[S=Q(q)-Et\]

Replacing in the H--J equation and considering the conditions:
\[I_{3}>I_{2}>I_{1}, \,\,\,\, 2EI_{1}<\lambda<2I_{3}E\]

We find the solution of (\ref{euler}):
\begin{eqnarray}
L_{1}&=&I^{1/2}_{1}\left(\frac{2EI_{3}-\lambda}{I_{3}-I_{2}}\right)cn\tau\nonumber\\
L_{2}&=&I^{1/2}_{2}\left(\frac{2EI_{3}-\lambda}{I_{3}-I_{2}}\right)sn\tau \label{landausol}\\
L_{3}&=&I^{1/2}_{3}\left(\frac{\lambda -2EI_{1}}{I_{3}-I_{1}}\right)dn\tau\nonumber\\
\tau&=&\sqrt{\frac{(\lambda-2EI_{1})(I_{3}-I_{2})}{I_{1}I_{2}I_{3}}}(t-t_{0})\nonumber
\end{eqnarray}

Where $sn\tau$, $cn\tau$, and $dn\tau$ are the Jacobi functions.
Note that (\ref{landausol}) is agree with the solution obtained by L. D. Landau in \cite{landau}.

Another Poisson structure can be obtained taking a Hamiltonian
$H=\frac{1}{2}C_{2}$, a Casimir $\Psi=C_{1}$ and a Poisson matrix:

\[ J^{(2)ab}=\varepsilon^{abc}\frac{\partial \Psi}{\partial X^{c}}
\]
 \begin{equation}\|J^{(2)ab}\|=
\left ( \begin{array}{lcr}
0&\frac{L_{3}}{I_{3}}&-\frac{L_{2}}{I_{2}}\\
-\frac{L_{3}}{I_{3}}&0&\frac{L_{1}}{I_{1}}\\
\frac{L_{2}}{I_{2}}&-\frac{L_{1}}{I_{1}}&0
\end{array}\right )
\label{euler2}
\end{equation}

The phase space of (\ref{euler}) with this structure, is a Poisson manifold
with the brackets:
 \begin{eqnarray}
\{L_{1},L_{2}\}_{M}&=&\frac{L_{3}}{I_{3}\nonumber} \\
\{L_{2},L_{3}\}_{M}&=&\frac{L_{1}}{I_{1}}\label{bracket2}\\
\{L_{3},L_{1}\}_{M}&=&\frac{L_{2}}{I_{2}}\nonumber
\end{eqnarray}

Using the Poisson map $\varphi:M\rightarrow N$, defined by:
\begin{eqnarray}
L_{1}&=&\sqrt{\frac{\lambda -p^2}{I_{2}I_{3}}}\cos q\nonumber\\
L_{2}&=&\sqrt{\frac{\lambda -p^2}{I_{1}I_{2}}}\sin q\label{eulerpoi2}\\
L_{3}&=&\frac{p}{\sqrt{I_{1}I_{2}}} \nonumber
\end{eqnarray}

the phase space $M$ is reduced to the submanifold $N$, defined by the equation: $\Psi
(L_{1},L_{2},L_{3})=C_{1}$. $N$ is symplectic in the variables $p$ and
$q$. With the Poisson map (\ref{eulerpoi2}) we obtain the canonical brackets
for $p$ and $q$
from (\ref{bracket2}).

Introducing the action by the transformation:
$q\rightarrow q$ and $p\rightarrow\frac{\partial S}{\partial q}$.
We have the H--J equation:
\begin{equation}
\frac{\partial S}{\partial t}+\frac{1}{2I_{2}I_{3}}\left[\lambda-\left(\frac{\partial
S}{\partial q}\right)^2\right]\cos^{2}q+\frac{1}{2I_{1}I_{3}}\left[\lambda-\left(\frac{\partial
S}{\partial q}\right)^2\right]\sin^{2}q+\frac{1}{2I_{1}I_{2}}\left(\frac{\partial
S}{\partial q}\right)^2=0
\label{HJeuler1}
\end{equation}

Taking the solution of (\ref{HJeuler1}) in the form:
\[S=Q(q)-Et\]

Note that in this additive separation of the variables, the constant parameter
$E$ is not the usual energy.

Replacing in the H--J equation (\ref{HJeuler1}) and considering the conditions:
\[I_{3}>I_{2}>I_{1}, \,\,\, 2EI_{1}I_{2}<\lambda<2EI_{3}I_{2}, \,\,\, \lambda=2I_{1}I_{2}I_{3}C_{1}\]

We find the solution of (\ref{euler}):
\begin{eqnarray}
L_{1}&=&\frac{1}{I^{1/2}_{2}}\left(\frac{\lambda-2EI_{1}I_{2}}{I_{3}-I_{1}}\right)cn\tilde{\tau}\nonumber\\
L_{2}&=&\frac{1}{I^{1/2}_{1}}\left(\frac{\lambda-2EI_{1}I_{2}}{I_{3}-I_{2}}\right)sn\tilde{\tau} \label{otrasol}\\
L_{3}&=&\frac{1}{I^{1/2}_{2}}\left(\frac{2EI_{2}I_{3}-\lambda}{I_{3}-I_{1}}\right)dn\tilde{\tau}\nonumber\\
\tilde{\tau}&=&\frac{1}{I_{1}I_{2}I_{3}}\sqrt{I_{1}(I_{3}-I_{2})(2EI_{2}I_{3}-\lambda)}(t-\tilde{t_{0}})\nonumber
\end{eqnarray}

It is a straightforward matter to realize that (\ref{otrasol})
is a solution of (\ref{euler}) and become to (\ref{landausol}) taking:
\[\lambda=2I_{1}I_{2}I_{3}C_{1}, \,\,\, 2E=C_{2}\]

\section{Conclusions}
In section II we have been able to construct two different structures based on
one symmetry vector $\eta$ and two conserved quantities $C_{1}$ and $C_{2}$
for one dynamical system. The Poisson structures $J^{(1)}$ and $J^{(2)}$ are built
as the antisymmetric product of the evolution vector $f$ and the symmetry
vector $\eta$ normalized using the deformation $K$ and $Q$ of $C_{1}$ and
$C_{2}$ respectively.

In section III we showed that the stability criterion involving the derivative
of a Casimir function parametrized by a Lagrange multiplier strongly depends
of the used Poisson structure. The situation is even worse, because we can
construct infinitely many Poisson structures, using the Poisson matrix defined
by:
\[J^{ab}=\mu\varepsilon^{abc}\frac{\partial D}{\partial X^{c}}\]

 Where
$D=D(C_{1},C_{2})$ is an arbitrary function of $C_{1}$ and $C_{2}$, and a
Hamiltonian $H=H(C_{1},C_{2})$, which is a function independent of $D$. We
have many Poisson structures depending on the different choises of arbritary
functions $D$ and $H$.(See for details \cite{little,hojman2}).

In the section IV, the Hamilton--Jacobi equations for a number of simple classical
systems were obtained using non--standard Hamiltonian, constructed with the
Hojman's techniques, developed in \cite{hojman1}--\cite{hojman7}.

 It is important to note that in the reduction of the phase
space $M$ of a dynamical system, the different values of the Casimir functions
define a foliation of $M$. Each $sheet$, specified by the values of the
casimir contains entire trajectories. i. e. solution of the equation of
motion, and cannot be connected to any other sheet, only states within a sheet
are transformable into one another. In other words each selection of the
Poisson structure and then each sheet yield a {\em different physical system},
though they all share the same equation of motion. Further, since classical
 H--J theory forms a link with quantum mechanics it is
of particular importance to examine alternative appoaches of constructing H--J
equation for one classical system, to see if they lead to a razonable quantum
mechanical systems.

Alternative H--J equations for Euler top are explicitly exhibited and solved.
We demonstrated that both obtained H--J equation yield the correct solutions
for the equation of motion, thought the $actions$ must be different.

 More examples will be discussed in forthcoming articles.

\section{Acknowledgments}

M.H. is deeply grateful to Instituto de Cooperaci\'on
Iberoamericana for the MUTIS Doctoral Fellowship, and to
Universidad de Chile for partial financial support (Proyecto 073).


\begin{thebibliography}{shepherd}

\bibitem{little} R.G. Littlejohn, A.I.P. Conf. Proc.{\bf 28}, 47
(1982).
\bibitem{hojman1} S.A. Hojman,"The construction of a Poisson structure out of
a symmetry and conservation law of a dynamical system", J. of Phys. A: Math. Gen. {\bf29}, 667
(1996).
\bibitem{hojman2} S.A. Hojman, "Construction of hamiltonian structure for
dynamical system from scratch" in "Instabilities and Non--Equilibrium Structures
V",
 Kluwer Academic Publishers, edited by E. Tirapegui and W. Zeller, 281 (1996).
\bibitem{hojman3}A. Gomberoff, S. Hojman "Non--standard construction of
Hamiltonian structures",J. Phys. A: Math. Gen. {\bf 30} (1997), 5077--5084.
\bibitem{hojman4} S. Hojman, et. al. "Minisuperspace example of
non--lagrangian quantization", Phys. Rev. D{\bf 45}, (1992), 3523--3527.
\bibitem{hojman5} M.P. Ryan Jr. and S. Hojman, ``Directions in General
Relativity'', Proceedings of the 1993 International Simposium, Maryland,
Papers
in Honor of Charles Misner, Cambridge University Press, edited by B.L. Hu,
M.P. Ryan Jr. and C.V.  Vishveshwara, \underline{1}, 300 (1993).
\bibitem{hojman6} S.A. Hojman, D. N\'u\~nez and M.P. Ryan Jr. in
``Proceedings
of the Sixth Marcel Grossmann Meeting'' (Part A), Kyoto, Jap\'on, H. Sato
and
T. Nakamura Editors, page 718 (World Scientific Publishing Co.,
Singapore, 1992).
\bibitem{hojman7} S.A. Hojman, D. N\'u\~nez, and M.P. Ryan, Jr., Phys. Rev. D
{\bf 45}, 3523 (1992).
\bibitem{wein} S. Weinberg, {\it Gravitation and Cosmology} (Wiley, New
York,
1975).
\bibitem{mukun2}E. C. G. Sudarshan, N. Mukunda et. al. "Evolution, symmetry, and canonical
structure in dynamics". Phys. Rev. D{\bf 23}, (1981)2189--2200.
\bibitem{mukun1}E. C. G. Sudarshan, N. Mukunda, "Classical Dynamics:A Modern
Perspective", Jhon Wiley \& Sons (1974).
\bibitem{marsd}D. D. Holm, J. E. Marsden, T. Ratiu, A. Weinstein, "Nonlinear
stability of fluid and plasma equilibria", Phys. Reports {\bf 123}, N 1,2 (1985).
\bibitem{marsd1}J. Marsden, A. Weistein, "Coadjoin orbits, vortices and
Clebsch variables for imcompressible fluids", Physica D{\bf 7}, (1983), 305--323.
\bibitem{holm}D. D. Holm, B. A. Kupershmidt "Noncanonical formulation of ideal
magnetohydrodynamics", Physica D{\bf 7}, (1983), 330--333.
\bibitem{kuper}B. A. Kupershmidt "On dual spaces of differntial Lie algebras"
Physica D{\bf 7}, (1983), 334--337.
\bibitem{bridges}T. J. Bridges "Poisson structure of the reversible 1:1
resonance", Physica D{\bf 112}, (1998), 40--49.
\bibitem{holmes1}D. Armbruster, J. Guckenheimer, P. Holmes, ''Heteroclinic
cycles and modulated travelling waves in systems with {\large O}(2)
symmetry'', Physica D{\bf 29}, (1988,) 257--282.
\bibitem{holmes2}P. Holmes et. al. ''Low--dimensional models of coherent
structures in turbulence'', Phys. Reports {\bf 287}, N4, (1997).
\bibitem{landau}L. D. Landau, E. M. Lifshitz "Mec\'anica", Editorial
Revert\'e, S. A. (1970).
\end{thebibliography}
\end{document}